\documentclass[aps,onecolumn,showpacs,nofootinbib,showkeys]{revtex4-2}
\usepackage[margin=3.0cm]{geometry}
\RequirePackage[T1]{fontenc}

\usepackage{epsfig,graphicx,amsmath,amssymb,bm}
\RequirePackage{mathptmx}      
\usepackage{mathrsfs}
\usepackage{amssymb}
\RequirePackage{color}
\usepackage{changes}
\usepackage{slashed}
\usepackage{multirow}
\usepackage{scalerel}
\usepackage{tikz-feynman}
\usepackage{textcomp}
\usepackage{subcaption}
\usepackage[english]{babel}
\usepackage{float}
\RequirePackage{hyperref}
\hypersetup{
    linktocpage,
    colorlinks,
    citecolor=purple,
    filecolor=black,
    linkcolor=blue,
    urlcolor=blue,
}
\usepackage{nccmath}

\begin{document}

\title{
Two-body decays of radially excited $\eta(1295)$ and $\eta(1475)$ mesons in the extended NJL model
}

\author{M.K. Volkov$^{1}$}\email{volkov@theor.jinr.ru}
\author{A.A. Pivovarov$^{1}$}\email{pivovarov@theor.jinr.ru}
\author{K. Nurlan$^{1,2}$}\email{nurlan@theor.jinr.ru}

\affiliation{$^1$ Bogoliubov Laboratory of Theoretical Physics, JINR, 141980 Dubna, Russia \\
            $^2$ Institute of Nuclear Physics, 050032, Almaty, Kazakhstan}   


\begin{abstract}
In the extended NJL model, two-body strong decays of the radially excited $\eta(1295)$ and $\eta(1475)$ mesons are described. It is shown that the two-body decays $\eta(1475)\to K_0^*K, a_0\pi$ play a dominant role in determining the width of the radially excited meson $\eta(1475)$. At the same time, the decays $\eta(1475)\to K^*K$ make a significantly smaller contribution to the total width.
It is shown that two-particle decays give only a negligible contribution to the total width of the $\eta(1295)$.
\end{abstract}


\maketitle
\section{INTRODUCTION}
Presently, $\eta$ mesons are actively investigated from both experimental and theoretical points of view~\cite{E852:2001ote,BaBar:2008rth,BESIII:2022chl,SND:2024qaq,Wu:2012pg,Achasov:2021yis,Gan:2020aco,Nakamura:2022rdd}. However, currently, there are no solid experimental data on the decays of the radially excited mesons $\eta(1295)$ and $\eta(1475)$. That is why, making a prediction for the most characteristic decays of these mesons in well-known models is of great interest. It is natural to refer to such phenomenological model as the Nambu--Jona-Lasinio (NJL) model \cite{Nambu:1961tp,Eguchi:1976iz,Ebert:1982pk,Volkov:1984kq,Volkov:1986zb,Volkov:1986zb,Ebert:1985kz,Vogl:1991qt,Klevansky:1992qe,Ebert:1994mf}, specifically to its extended version allowing one to describe both the ground and first radially excited meson states~\cite{Volkov:1996br,Volkov:1996fk,Volkov:1999yi,Volkov:2005kw,Volkov:2017arr}. In the framework of this model, numerous strong, weak and electromagnetic meson interactions at low energies are described~\cite{Volkov:1999yi,Volkov:2017arr}.

The additional difficulty is in describing the mixing of $\eta$ mesons, since even in the ground states, there is a singlet-octet mixing caused by the gluon anomaly. When taking into account the first radially excited states, it is necessary to consider the mixing of the four $\eta$ mesons. This was carried out in the work~\cite{Volkov:1999xf} and successfully applied to describe the mass spectrum of the $\eta$ mesons.

The two-body anomalous electromagnetic decays of the mesons $\eta(1295)$ and $\eta(1475)$ were described in the framework of the NJL model~\cite{Vishneva:2013mga}. However, these decays make an insignificant contribution to the total decay width.
In the energy region above 1.4 GeV, pseudoscalar states $\eta(1405)
$ and $\eta(1475)$ are observed \cite{ParticleDataGroup:2024cfk}. To explain the nature of these mesons, attempts were made to consider the first state as a glueball candidate and the second state as the first radial excitation of the $\eta'$ meson \cite{Gutsche:2009jh,Li:2009rk}. However, this contradicts the results of calculations in lattice QCD, which predicts a pseudoscalar glueball mass in the range \(2.4\)-\(2.6\)~GeV \cite{Richards:2010ck,Dudek:2013yja}. The quark NJL model predicts only one state, a radially excited $\eta(1475)$, above 1.4 GeV with mass $M_{\eta(1475)}=1470$ MeV \cite{Volkov:1999yi,Volkov:1999xf} which is in agreement with experiment $M^{exp.}_{\eta(1475)}=1476 \pm 4$ MeV \cite{ParticleDataGroup:2024cfk}. Therefore, only this state will be considered in further calculations.

In the present work, the strong two particle decays of the states $\eta(1295)$ and $\eta(1475)$ are considered. As it is shown below, these decays play the main role in the case of the meson $\eta(1475)$ and almost determine its total width. At the same time, in the case of the meson $\eta(1295)$, these decays give a less noticeable contribution. It indicates that three-body decays give a dominant contribution to the full width of this meson. The two-particle decays studied in this work can be used as intermediate states in further research of three particle decays of these mesons.
\section{EXTENDED NJL MODEL}
To describe the two-particle decays of the excited mesons $\eta(1295)$ and $\eta(1475)$, we will use the extended $U(3)\times U(3)$ NJL model. It is a chiral quark model that successfully describes the interactions and mass spectra of the ground and first radially excited states of scalar, pseudoscalar, vector and axial-vector meson nonets using a small number of parameters that are fixed in the process of constructing the model ~\cite{Volkov:1996br,Volkov:1996fk,Volkov:1999yi,Volkov:2005kw,Volkov:2017arr}. A distinctive feature of this version of the quark NJL model is a rather large value of the cutoff parameter, namely $\Lambda_{4} = 1260$~MeV. This provides a basis for attempts to include mesons in the first radially excited states in the model with partial conservation of chiral symmetry. Such a model was constructed in ~\cite{Volkov:1996br,Volkov:1996fk}. Radially excited states were described using a simple form factor in the form of a second-degree polynomial in the relative transverse momentum of quarks in the meson. The numerical coefficient of $k_{\perp}^2$ (the slope parameter $d$) in the form factor can be fixed without using experimental data, based on the requirement that the quark condensate remains unchanged after including radially excited states. This leads to the preservation of the values of the basic parameters of the original standard model such as the quark masses and the ultraviolet cutoff parameter. Mixing of mesons in radially excited states with their ground states are taken into account. This leads to the appearance of non-diagonal terms in the free Lagrangian, reflecting possible transitions between the ground and excited states. The diagonalization of the free Lagrangian is achieved by introducing mixing angles or an appropriate matrix, in particular for $\eta$ mesons. As a result, when describing interactions in the extended model, the number of arbitrary parameters does not exceed that of the standard NJL model.

The quark-meson Lagrangian for the strong interaction of meson fields necessary for describing the processes considered here, in the NJL model takes the form \cite{Volkov:1999xf, Volkov:1999yi, Volkov:2005kw,Volkov:2017arr}
\begin{eqnarray}
{\cal L}_{int} = \bar{q} \biggl[ 
i \gamma_5 \sum_{i = \pm, 0} \lambda^\pi_i (A_\pi\pi^i + B_\pi\pi'^{i})+ i \gamma_5 \sum_{i = \pm, 0} \lambda^K_i (A_K K^i + B_K K'^{i})
\\ \nonumber
+ \sum_{i = \pm,0} \lambda^{a_0}_i (A_{a_0}a^i_0 + B_{a_0}a'^i_0)
+ \frac{1}{2} \gamma_\mu \sum_{i = \pm,0} \lambda^K_i (A_{K^*}K^*_i +B_{K^*}{K^*}'_i)
\\ \nonumber
+ \sum_{i = u,s} \lambda_i (A^i_{f_0} f_0 + B^i_{f_0} f'_0)
+\sum_{i = \pm,0} \lambda^K_i (A_{K^*_0}K^{*i}_0 + B_{K^*_0}{K^*_0}'^i)
\\ \nonumber
+ i\gamma^{5} \sum_{i = u, s} \lambda_{i} \left[A^{i}_{\eta}\eta + A^{i}_{\eta'}\eta' + A^{i}_{\hat{\eta}}\hat{\eta} + A^{i}_{\hat{\eta}'}\hat{\eta}'\right] 
\biggl]q,
\end{eqnarray}
where $q$ and $\bar{q}$ are u, d and s quark fields with constituent quark masses $m_{u} \approx m_{d} = 270$~MeV, $m_{s} = 420$~MeV; $f_0=f_0(500)$, the $\eta'$ meson corresponds to the physical state $\eta'(958)$ and the $\hat{\eta} = \eta(1295)$, $\hat{\eta}' = \eta(1475)$$\hat{\eta}$, $\hat{\eta}'$ mesons correspond to the first radial excitation mesons $\eta$ and $\eta'$; 
$\lambda$ are linear combinations of the Gell-Mann matrices \cite{Volkov:1999yi, Volkov:2017arr}
\begin{eqnarray}
\label{verteces1}
	A_{M} = A^0_M \left[g_{M}\sin(\theta_M + \theta^0_M) +
	g'_{M}f_{M}(k_{\perp}^{2})\sin(\theta_M - \theta^0_M) \right], 
    \nonumber\\
	B_{M} = - A^0_M \left[g_{M}\cos(\theta_M + \theta^0_M) +
	g'_{M}f_{M}(k_{\perp}^{2})\cos(\theta_M + \theta^0_M) \right],
\end{eqnarray}
where $A^0_{M} = 1/{\sin(2\theta_{M}^{0})}$. The subscript M indicates the corresponding meson; The factors $A^u_{f_0}$ and $A^s_{f_0}$ for the scalar state $f_0(500)$ have the form
\begin{eqnarray}
    \label{verteces2}
    A^{u}_{f_0} & = & -0.98 g_{f^u_0} + 0.02 g'_{f^u_0} f_{uu}(k_{\perp}^{2}), \nonumber\\
    A^{s}_{f_0} & = & 0.27 g_{f^s_0} - 0.03 g'_{f^s_0} f_{ss}(k_{\perp}^{2}).
\end{eqnarray}

 The values of the mixing angles are given in Table~\ref{tab_mixing}.

\begin{table}[htbp]
\caption{Values of the mixing angles of the ground and first radially excited mesons}
\begin{center}
\small
\begin{tabular}{lcccccl}
\hline
\hline
   & $\pi$ & $K$ & $a_0$ & $K^*_0$ & $K^*$ \\
\hline
$\theta_M$	& $59.48^{\circ}$	&  $58.11^{\circ}$  & $72.0^{\circ}$ & $74.0^{\circ}$ & $84.7^{\circ}$ \\
$\theta^0_M$	& $59.12^{\circ}$	& $55.52^{\circ}$  & $61.50^{\circ}$ & $60.0^{\circ}$ & $59.14^{\circ}$ \\
\hline
\hline
\end{tabular}
\label{tab_mixing}
\end{center}
\end{table}

The mixing angles for K and $\pi$ mesons $\theta \approx \theta_0$, so for the ground states of these mesons one can use $A_\pi = g_\pi$ and $A_K=g_K$. 

For the $\eta$ mesons, the factor $A$ takes a slightly different form. This is due to the fact that in the case of the $\eta$ mesons, four states are mixed:
    \begin{eqnarray}
    \label{verteces3}
    A^{u}_{M} & = & g_{\eta^{u}} a^{u}_{1M} + g'_{\eta^{u}} a^{u}_{2M} f_{uu}(k_{\perp}^{2}), \nonumber\\
    A^{s}_{M} & = & g_{\eta^{s}} a^{s}_{1M} + g'_{\eta^{s}} a^{s}_{2M} f_{ss}(k_{\perp}^{2}).
    \end{eqnarray}
        
Here $f(k_\perp^2) = (1 + d k_\perp^2) \Theta(\Lambda^2 - k_\perp^2)$ is the form factor describing the first radially excited meson states. The slope parameters $d_{uu} = -1.784 \times 10^{-6}$~MeV$^{-2}$ and $d_{ss} = -1.737 \times 10^{-6}$~MeV$^{-2}$ are unambiguously fixed from the condition of constancy of the quark condensate after the inclusion of radially excited states and depend only on the quark composition of the corresponding meson \cite{Volkov:1999yi, Volkov:2005kw,Volkov:2017arr}. The values of the mixing parameters $A$ are shown in Table II.
\begin{table}[h!]
\begin{center}
\begin{tabular}{ccccc}
\hline
\hline
   & $\eta$ & $\eta(1295)$ & $\eta'$ & $\eta(1475)$ \\
\hline
$a^{u}_{1}$		& 0.71			& 0.62            &-0.32             & 0.56    \\
$a^{u}_{2}$		& 0.11			& -0.87           & -0.48            & -0.54   \\
$a^{s}_{1}$               & 0.62                        & 0.19            & 0.56             & -0.67 \\
$a^{s}_{2}$               & 0.06                       & -0.66           & 0.3               & 0.82 \\
\hline
\hline
\end{tabular}
\end{center}
\caption{Mixing parameters of $\eta$ mesons \cite{Volkov:1999yi}.}
\label{tab_eta}
\end{table}   
 
The quark-meson coupling constants have the form
\begin{eqnarray}
\label{Couplings}
   g_{\pi} = g_{\eta^{u}}=\left(\frac{4}{Z_{\pi}}I_{20}\right)^{-1/2}, 
\, g'_{\pi}=g'_{\eta^{u}} =  \left(4 I_{20}^{f^{2}}\right)^{-1/2}, 
\, g_{\eta^{s}}=\left(\frac{4}{Z_{\eta_s}}I_{02}\right)^{-1/2},
\nonumber\\
\, g'_{\eta^{s}} = \left(4 I_{02}^{f^{2}}\right)^{-1/2},
\,   g_{f^u_0} = g_{a_0} = \left(4I_{20}\right)^{-1/2}, 
\, g_{f^s_0} = \left(4I_{02}\right)^{-1/2},
\, g'_{f^u_0} = g'_{a_0} = \left(4I_{20}^{f^{2}}\right)^{-1/2}, 
\nonumber\\
\, g'_{f^s_0} = \left(4I_{02}^{f^{2}}\right)^{-1/2},
\,   g_{K} =\left(\frac{4}{Z_K}I_{11}\right)^{-1/2}, 
\, g'_{K} =\left(4I^{f^2}_{11}\right)^{-1/2},
\, g_{K^*_0} =\left(4I_{11}\right)^{-1/2}, 
\nonumber\\
\, g'_{K^*_0} =\left(4I_{11}^{f^{2}}\right)^{-1/2}, 
\, g_{K^*} =\left(\frac{2}{3}I_{11}\right)^{-1/2},\,
\, g'_{K^*} =\left(\frac{2}{3}I_{11}^{f^{2}}\right)^{-1/2}, 
\end{eqnarray}
where $Z_\pi$ and $Z_{\eta s}$ are additional renormalization constants appearing in pseudoscalar and axial-vector transitions \cite{Volkov:2005kw}.

Divergent integrals appearing in the quark loops are
\begin{eqnarray}
	I_{n_{1}n_{2}}^{f^{m}} =
	-i\frac{N_{c}}{(2\pi)^{4}}\int\frac{f^{m}(k^2_{\perp})}{(m_{u}^{2} - k^2)^{n_{1}}(m_{s}^{2} - k^2)^{n_{2}}}\Theta(\Lambda_{3}^{2} - k^2_{\perp})
	\mathrm{d}^{4}k,
\end{eqnarray}
where $\Lambda_3=1030$ MeV is the three-dimensional cutoff parameter, the value of the four-dimensional cutoff parameter is $\Lambda_4=1260$ MeV \cite{Volkov:2005kw}. 

\section{Amplitudes and decay widths} 

The diagram describing the decay $\eta(1475) \to K_0^* K$ is shown in Fig.~1.

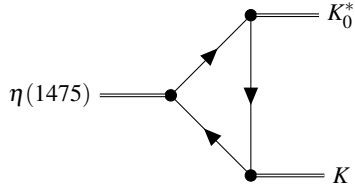
\begin{figure*}[t]
   \centering
    \begin{tikzpicture}
     \begin{feynman}
      \vertex (a) {\(\eta(1475) \)};
      \vertex [dot, right=1.6cm of a] (b) {};
      \vertex [dot, above right=1.5cm of b] (c) {};
      \vertex [dot, below right=1.5cm of b] (d) {};
      \vertex [right=1.2cm of c] (g) {\(K^*_0\)};
      \vertex [right=1.2cm of d] (f) {\(K \)};
      \diagram* {
        (a) -- [double] (b),
        (b) -- [fermion] (c),
        (c) -- [fermion] (d),
        (d) -- [fermion] (b),  
        (c) -- [double] (g),         
        (d) -- [double] (f),
      };
     \end{feynman}
    \end{tikzpicture}
   \caption{Triangle quark diagram for a two body decay with the scalar meson $\eta(1475) \to K^*_0 K$}
 \label{diagram1}
\end{figure*}%

The amplitude of this decay in the NJL model takes the following form:
\begin{eqnarray}
\mathcal{M}(\eta(1475) \to K^{*+}_0 K^-) = 8\left(m_s I_{11}^{K_0^*K\eta^u} - \sqrt{2}m_u I_{11}^{K_0^*K\eta^s }\right),
\end{eqnarray}
where the integrals over the quark triangle are
\begin{eqnarray}
\label{int_q}
I_{n_{1}n_{2}}^{M_1 M_2\dots} =
	-i\frac{N_{c}}{(2\pi)^{4}}\int\frac{A_{M_1}A_{M_2}\dots}{(m_{u}^{2} - k^2)^{n_{1}}(m_{s}^{2} - k^2)^{n_{2}}}\Theta(\Lambda_{3}^{2} - k^2_{\perp}),
	\mathrm{d}^{4}k.
\end{eqnarray}
Here $M$ denotes the corresponding meson, and $A_M$ are the coefficients from the Lagrangians~(\ref{verteces1}), (\ref{verteces2}), and (\ref{verteces3}).

The decay width can be calculated using the next formula 
\begin{eqnarray}
\label{width_formula}
\Gamma(\eta(1475) \to K^{*+}_0K^-) =
\frac{1}{2J_{\eta}+1} \frac{\sqrt{E^2_K - M^2_K}}{8\pi \, M^2_{\eta(1475)}} \cdot \, {\mid \mathcal{M} \mid}^2,
\end{eqnarray}
where $J_{\eta}=0$, $E_K = (M^2_{\eta(1475)}+M^2_{K}-M^2_{K^*_0})/2M^2_{\eta(1475)}$ is the energy of the $K$ meson in the rest system of the $\eta(1475)$ meson. The meson masses are taken from PDG \cite{ParticleDataGroup:2024cfk}. The uncertainty of numerical calculations within the model is estimated at $\pm15\%$ \cite{Volkov:2005kw,Volkov:2017arr}. Note that the main source affecting the accuracy of the model is the violation of the chiral symmetry, associated mainly with the non-zero current quark masses. This leads to partial non-conservation of the axial current (the PCAC principle). This estimate of the calculation accuracy also includes uncertainties in the meson mass values. As statistical analysis of numerous calculations show that in most cases satisfactory agreement with experiment at low energies in the model is achieved within the limits of the above-mentioned uncertainty.

The width of this decay is
\begin{eqnarray}
\Gamma(\eta(1475) \to K^{*+}_0 K^-) = 17.4 \pm 2.6 \textrm{ MeV}.
\end{eqnarray}
Then taking into account different charge distributions of mesons in the final states, the width of the decay is
\begin{eqnarray}
\Gamma(\eta(1475) \to K^{*}_0 K) = 69.6 \pm 10.4 \textrm{ MeV}.
\end{eqnarray}

The decay $\eta(1475) \to a_0^- \pi^+$ is described by the amplitude
\begin{equation}
\label{14a0pi}
\mathcal{M}(\eta(1475) \to a_0^- \pi^+) = 16 m_u I_{20}^{a_0 \pi \eta^u},
\end{equation}
where the loop integral $I_{20}^{a_0 \pi \eta^u}$ is defined in (8). Its width is
\begin{equation}
\Gamma(\eta(1475) \to a_0^- \pi^+) = 5.4 \pm 0.8 \textrm{ MeV}.
\end{equation}
The amplitudes of the decays $\eta(1475) \to a_0^+ \pi^-$, $a_0^0 \pi^0$ have a similar structure. As a result, for the width we obtain the estimate
\begin{equation}
\Gamma(\eta(1475) \to a_0 \pi) = 16.2 \pm 2.4 \textrm{ MeV}.
\end{equation}

The decay $\eta(1475) \to f_0 \eta$ includes triangle diagrams with $u$, $d$ and $s$ quark fields. The decay amplitude takes the form
\begin{equation}
\mathcal{M}(\eta(1475) \to f_0 \eta) = 16 \left( m_u I_{20}^{\hat{\eta}^u f_0^u \eta^u} - \sqrt{2} m_s I_{02}^{\hat{\eta}^s f_0^s \eta^s} \right),
\end{equation}
where $\hat{\eta}$ corresponds to the state $\eta(1475)$. For the decay width we obtain
\begin{equation}
\label{14f0eta}
\Gamma(\eta(1475) \to f_0 \eta) = 0.98 \pm 0.14 \, \textrm{MeV}.
\end{equation}

The amplitude of the decay involving the vector meson $\eta(1475) \to K^{*-} K^+$, in the NJL model takes the following form:
\begin{eqnarray}
&M(\eta(1475) \to K^{*-} K^+) = -2\left(I_{11}^{KK^*\eta^u} + \sqrt{2}I_{11}^{KK^*\eta^s}\right)& \nonumber\\
& \times\left[2\left(1 - \frac{m_s I_{11}^{K_1K^*\eta^u} + \sqrt{2}m_u I_{11}^{K_1K^*\eta^s}}{I_{11}^{KK^*\eta^u} + \sqrt{2}I_{11}^{KK^*\eta^s}}I_{11}^{K_1K}\frac{m_s + m_u}{M_{K_1}^2}\right)p_K + p_{K^*}\right]^\mu e_\mu(p_{K^*}),&
\end{eqnarray}
where $p_K$ and $p_{K^*}$ are the momenta of the final states, and $e_\mu(p_{K^*})$ is the polarization vector of the $K^*$ meson. The width of this process is
\begin{equation}
\Gamma(\eta(1475) \to K^{*-} K^+) = 0.59 \pm 0.09 \textrm{ MeV}.
\end{equation}
Then, taking into account different charge distributions, the width of the decay $\eta(1475) \to K^* K$ is
\begin{equation}
\Gamma(\eta(1475) \to K^* K) = 2.36 \pm 0.35 \textrm{ MeV}.
\end{equation}

There is the experimental estimation for this process~\cite{Edwards:1982nc}:
\begin{equation}
\frac{\Gamma(\eta(1475) \to KK^*)}{\Gamma(\eta(1475) \to KK^*) + \Gamma(\eta(1475) \to a_0(980)\pi)} < 0.25.
\end{equation}

Our theoretical estimation for this expression $\approx 0.13$, that is in agreement with the experimental data. The experimental value for the total width of the meson $\eta(1475)$ is $\Gamma_{\eta(1475)} = 96 \pm9$ MeV \cite{ParticleDataGroup:2024cfk}. As can be seen from the calculations, the total width of this meson is almost completely determined by the decays $\eta(1475) \to K_0^* K$ and $\eta(1475) \to K^* K$.

The partial widths of the decays $\eta(1295) \to a_0 \pi$ and $\eta(1295) \to f_0 \eta$ are obtained by replacing the vertices $\eta(1475) \to \eta(1295)$ in the quark-loop integrals in the amplitudes (\ref{14a0pi}) and (\ref{14f0eta}). The theoretical estimates obtained in the extended NJL model for these decays are
\begin{equation}
\Gamma(\eta(1295) \to a_0 \pi) = 2.33 \pm 0.35 \, \textrm{MeV}, \quad
\Gamma(\eta(1295) \to f_0 \eta) = 1.55 \pm 0.23 \, \textrm{MeV}.
\end{equation}

Since decays into $K_0^* K$ and $K^{*} K$ are impossible in the case of the $\eta(1295)$ meson, the contribution from two-particle decays to the total width of this meson turns out to be small.


\section{CONCLUSION}

In this work, the main two-body decay widths of the first radially excited states $\eta(1295)$ and $\eta(1475)$ have been calculated within the extended NJL model. As calculations have shown, the total width of the $\eta(1475)$ meson is largely determined by decays into strange states $K_0^* K$. Unfortunately, sufficiently accurate experimental data on these decays are not yet available.

In the case of the $\eta(1295)$ state, the two-particle decays into $a_0 \pi$ and $f_0 \eta$ considered in this work give only a small contribution to the total width. This suggests that the main contribution to the width of this meson should come from three-particle decays. At the same time, two-particle decays play an important role as intermediate states in three-particle decays.

There are currently no solid experimental data on two-body decays of radially excited states $\eta(1295)$ and $\eta(1475)$. Therefore, we consider the obtained results as predictions for future experiments. We note that numerous calculations of various processes involving first radially excited mesons have shown satisfactory agreement between previously obtained theoretical estimates and experimental data \cite{Volkov:1999yi, Volkov:2017arr}. In particular, with the participation of excited states $\eta(1295)$ and $\eta(1475)$, the widths of radiative decays $\eta \to V\gamma$ (where $V = \rho, \omega, \phi$) \cite{Vishneva:2013mga}, two-photon decays and photoproduction processes of $\eta$ mesons on electrons were described in \cite{Arbuzov:2011rb}, and the total cross sections for their production in $e^+ e^- \to \gamma[\eta, \eta', \eta(1295), \eta(1475)]$ annihilation in \cite{Ahmadov:2013ksa}. This gives hope for the reliability of the results obtained in this work and for their confirmation in future experimental studies at $e^+ e^-$ colliders (Belle II, BES III, Super c-tau factory) and high-intensity hadron factories.

 \subsection*{Acknowledgements}
The authors thank Prof. A. B. Arbuzov for useful discussions. 

\end{document}